\documentclass{iopjournal}

\usepackage{graphicx}
\usepackage{dcolumn}
\usepackage{bm}
\usepackage{xcolor}
\usepackage{amsmath, amssymb} 
\usepackage{hyperref}
\hypersetup{colorlinks=true, allcolors={blue},breaklinks=true}

\begin{document}

\articletype{Paper} 

\title{An improved reliability factor for quantitative low-energy electron diffraction}

\author{Alexander M. Imre$^1$, Lutz Hammer$^2$, Ulrike Diebold$^1$, Michele Riva$^1$, and Michael Schmid$^{1,*}$}

\affil{$^1$Institute of Applied Physics, TU Wien, Vienna, Austria}

\affil{$^2$Solid State Physics, Friedrich-Alexander-Universit{\"a}t Erlangen-N{\"u}rnberg, Erlangen, Germany}

\affil{$^*$Author to whom any correspondence should be addressed.}

\email{schmid@iap.tuwien.ac.at}


\begin{abstract}
Quantitative low-energy electron diffraction [LEED $I(V)$ or LEED $I(E)$, the evaluation of diffraction intensities $I$ as a function of the electron energy] is a versatile technique for the study of surface structures. The technique is based on optimizing the agreement between experimental and calculated intensities. Today, the most commonly used measure of agreement is Pendry's $R$ factor $R_\mathrm{P}$. While $R_\mathrm{P}$ has many advantages, it also has severe shortcomings, as it is a noisy target function for optimization and very sensitive to small offsets of the intensity. Furthermore, $R_\mathrm{P} = 0$, which is meant to imply perfect agreement between two $I(E)$ curves can also be achieved by qualitatively very different curves. We present a modified $R$ factor $R_\mathrm{S}$, which can be used as a direct replacement for $R_\mathrm{P}$, but avoids these shortcomings. We also demonstrate that $R_\mathrm{S}$ is as good as $R_\mathrm{P}$ or better in steering the optimization to the correct result in the case of imperfections of the experimental data, while another common $R$ factor,  $R_\mathrm{ZJ}$ (suggested by Zanazzi and Jona) is worse in this respect. 
\end{abstract}

%
%
\section{\label{sec:intro}Introduction}

Quantitative low-energy electron diffraction (LEED) analyzes the diffraction intensities $I$ as a function of the electron energy $E$ or equivalently, electron acceleration voltage $V$. Thus, the technique is also known known as LEED $I(V)$ or LEED $I(E)$, and the data are commonly referred to as $I(V)$ or $I(E)$ curves. Quantitative LEED is the oldest method for obtaining high-accuracy data in surface crystallography and has the advantage that it uses rather simple instrumentation (LEED optics), which is available in many ultrahigh-vacuum surface science systems \cite{pendry_book_1974,van_hove_leed_1986,van_hove_ssr_1993,heinz_review_1998,heinz_electron_2013, fauster_surface_2020,moritz_book_2022,held_susc_2025}.  The ViPErLEED project \cite{kraushofer_viperleed_2025,schmid_viperleed_2025,doerr_viperleed_2026,imre_jax_2026} aims at simplifying the use of this important method, and also includes further method development, as in the current work.

LEED $I(E)$ analysis is based on the comparison of the calculated $I(E)$ curves with the experimental ones.  The agreement between calculated and experimental $I(E)$ curves is described by a reliability factor ($R$ factor). Assuming that the structure model is qualitatively correct, variation of the structural parameters (such as atom coordinates and vibration amplitudes) to minimize the $R$ factor leads to the best-fit structure. Many different types of $R$ factors have been proposed and used for LEED $I(E)$ \cite{van_hove_leed_1986}. The simplest one is essentially an $L^2$ norm%
\begin{equation}
  R_2 = \frac{1}{\int{I_\mathrm{exp}^2\mathrm{d}E}}\int{(I_\mathrm{exp}-c I_\mathrm{th} )^2\mathrm{d}E}
  \label{eq:R2}
\end{equation}
where the subscripts ``exp'' and ``th'' refer to the experimental and calculated (theory) data, respectively. The factor $c=\int{I_\mathrm{exp}\mathrm{d}E}/\int{I_\mathrm{th}\mathrm{d}E}$ takes into account that experiments determine only relative intensities; the ratio between the outgoing electron current of a given diffraction maximum (named diffraction ``beam'' in LEED studies) and the primary current of the incident electrons is usually not known quantitatively. In most cases, $R_2$ is not an ideal metric for LEED $I(E)$ analysis, however. If two $I(E)$ curves are qualitatively very similar, but the relative peak heights differ, $R_2$ values can be rather poor. Different relative peak heights between experiment and calculation are very common, however, e.g., because of energy-dependent gain factors in experiment or an inaccurate description of energy-dependent inelastic scattering in the calculations. Furthermore, $R_2$ is not very sensitive to the positions of the minima of the $I(E)$ curves. This is undesirable since the minima carry important information and are less susceptible to various imperfections of experimental $I(E)$ curves that strongly influence the maxima (e.g., uneven transmission of the grids, shot noise).

A more elaborate $R$ factor was proposed by Zanazzi and Jona \cite{zanazzi_susc_1977}. It is based on the first and second derivatives of the $I(E)$ curves:
\begin{equation}
  R_\mathrm{ZJ} \propto \int{\frac{|I'_\mathrm{exp}-c I'_\mathrm{th}|\times |I''_\mathrm{exp}-c I''_\mathrm{th}|}{|I'_\mathrm{exp}|+\max(|I'_\mathrm{exp}|)}\mathrm{d}E}
  \label{eq:RZJ}
\end{equation}
The denominator is meant to put increased weight on regions of minima and maxima, where the first derivative $I'$ vanishes.%
\footnote{In practice, the denominator in the integral of $R_\mathrm{ZJ}$, $|I'_\mathrm{exp}|+\max(|I'_\mathrm{exp}|)$ substantially exceeds its minimum value of $\max(|I'_\mathrm{exp}|)$ only at the slopes of high peaks and only the influence of these gets reduced by a factor that can reach $1/2$. For weak peaks, $|I'_\mathrm{exp}|$ is usually much less than $\max(|I'_\mathrm{exp}|)$. Therefore, the denominator slightly reduces the weights for high-intensity peaks, but to a far lesser degree than Pendry's $R$ factor, where the absolute height of a maximum does not affect its influence on the $R$ factor.}
Like $R_2$, $R_\mathrm{ZJ}$  uses an energy-independent scale factor $c$ between experimental and calculated intensities. This makes it sensitive to slow variations with the energy of the experimental or calculated intensity scale factor. A further disadvantage of $R_\mathrm{ZJ}$ lies in the fact that it heavily depends on the second derivative $I''$. Experimental $I(E)$ curves inevitably contain some noise, but also the calculated curves often exhibit numerical noise (e.g., due to matrix inversions, if not all matrices are sufficiently well-conditioned to avoid numerical inaccuracies). Since the derivatives are calculated numerically, the second derivative $I''$ is more sensitive to noise and small numerical errors than $I'$.

Pendry's $R$ factor $R_\mathrm{P}$ avoids these issues since it is based on the logarithmic derivative $L = \mathrm{d}\ln I/\mathrm{d}E$ \cite{pendry_jphysc_1980}. A direct comparison of the $L$ functions of the experimental and theoretical $I(E)$ curves is not feasible because $L$ approaches infinity if $I$ approaches zero.%
\footnote{When expressing $L$ as $I'/I$, both the numerator and the denominator reach zero at a minimum of intensity $I=0$. Since $I(E)$ and its derivatives are smooth functions, one will typically have $I\propto (E-E_0)^2$ and $I'\propto E-E_0$, with $E_0$ being the energy of the minimum. This leads to $L \propto (E-E_0)^{-1}$, which diverges at the minimum (for $E \to E_0$).}
 Therefore, $R_\mathrm{P}$ does not use $L$ directly but rather compares the experimental  and calculated $Y_\mathrm{P}$ functions (named $Y_\mathrm{exp}$ and $Y_\mathrm{th}$, respectively):
\begin{align}
  L &= \frac{\mathrm{d}\ln I}{\mathrm{d}E} = \frac{I'}{I} \label{eq:L}\\
  Y_\mathrm{P} &= \frac{L^{-1}}{L^{-2}+V_\mathrm{0i}^2} = \frac{L}{1+V_\mathrm{0i}^2 L^2} = \frac{I \times  I'}{I^2+V_\mathrm{0i}^2 I'^2}
\label{eq:YPe}\\[2pt]
  R_\mathrm{P} &= \frac{ \int{(Y_\mathrm{exp}-Y_\mathrm{th})^2 \,\mathrm{d}E}}
    {\int{ (Y_\mathrm{exp}^2 + Y_\mathrm{th}^2)\,\mathrm{d}E}}
  \label{eq:RPe}
\end{align}
Note that the $Y_\mathrm{P}$ function approaches zero for both $L\to 0$ and $L \to \pm\infty$. This means that $R_\mathrm{P}$ is insensitive to the high (or even infinite) values that the logarithmic derivative $L$ reaches in minima with the intensity approaching 0.

$V_\mathrm{0i}$ is the imaginary part of the inner potential%
\footnote{LEED $I(E)$ calculations are based on the concept of muffin-tin potentials and replace the atoms by point scatterers; the inner potential is the potential between the scatterers and determines the propagation of the waves in the solid.  The imaginary part $V_\mathrm{0i}$ of the inner potential leads to an exponential decay and is also known as optical potential \cite{pendry_book_1974}.}
and describes the inelastic scattering of the electrons. It provides a natural energy scale: Assuming single scattering (kinematic diffraction theory), the peaks of $I(E)$ are Lorentzians with a full width at half maximum of $2|V_\mathrm{0i}|$ \cite{pendry_jphysc_1980}. LEED must be described by multiple scattering (dynamic scattering theory), however. For typical $I(E)$ curves it turns out that the minimum distance between two maxima (or two minima) is about $2|V_\mathrm{0i}|$; pronounced maxima with a deep minimum in between are usually separated by ${\approx}\hspace{0.2ex}3|V_\mathrm{0i}|$ or more. Typical values of $V_\mathrm{0i}$ are between $-3.5$ and $-6$\,eV.

Pendry's $R$ factor takes values between 0 and 2: A value of 0 is achieved when the experimental and calculated $I(E)$ functions are identical (ignoring a constant scale factor), and $R_\mathrm{P} = 1$ indicates uncorrelated curves. Values above 1 occur for anti-correlation between the $Y$ functions (usually related to anti-correlation of the $I(E)$ curves). Depending on the type of system, $R_\mathrm{P} \lesssim 0.2$ is usually considered good enough to lend high credibility to a structural model; values below 0.1 are considered excellent  \cite{heinz_electron_2013,moritz_book_2022}.

Apart from its insensitivity to slow variations of the scale factor between experiment and calculation, and a high sensitivity to the positions of the minima and maxima, a further advantage of $R_\mathrm{P}$ is the fact that it is based only on a local (point-by point) comparison of the $I(E)$ curves. Since the $Y$ function oscillates rapidly between its two extreme values $\pm 1/|2V_\mathrm{0i}|$ (see section \ref{sec:similar} for an example), the denominator of equation (\ref{eq:RPe}) is roughly proportional to the energy span of the integral.%
\footnote{In a LEED analysis, usually many beams are compared and an overall $R$ factor is determined. In this case, the integrals in equation (\ref{eq:RPe}) are replaced by the sums of the integrals for the individual beams. Alternatively, one could also take a sum of the $R$ factors of the individual beams weighted with their energy spans. In practice, the difference between these two approaches is very small: The average of $Y_\mathrm{P}^2$ is about $0.1/|V_\mathrm{0i}^2|$ to $0.15/|V_\mathrm{0i}^2|$ for all curves (depending on the system). For reasonably low $R_\mathrm{P}$, $Y_\mathrm{exp}$ and $Y_\mathrm{th}$ are well correlated; then  the denominator of equation (\ref{eq:RPe}) is about $0.4/|V_\mathrm{0i}^2|$ to $0.6/|V_\mathrm{0i}^2|$ times the energy span.}
While $R_2$ and $R_\mathrm{ZJ}$ are dominated by the highest peaks, for $R_\mathrm{P}$ the agreement of two curves in a particular region will always have the same effect on $R_\mathrm{P}$, largely independent of the remaining regions of the $I(E)$ curve(s). These advantages of $R_\mathrm{P}$ are not merely theoretical; based on real-world experimental data it has been shown that minimizing $R_\mathrm{P}$ in a structure search yields more accurate results than using $R_2$ \cite{sporn_accuracy_1998}.

Finally, $R_\mathrm{P}$ does not only provide a measure for the level of agreement between two $I(E)$ curves (or two sets of such curves), it also forms the basis for an estimate of the errors of a LEED $I(E)$ analysis. The confidence region of the fit parameters obtained by minimizing $R_\mathrm{P}$ roughly reaches to the point where the $R$ factor calculated with the modified parameter (or parameter combination in the case of strong coupling between parameters \cite{blum_susc_2001}) reaches the value $R_\mathrm{P,min} + \mathrm{var} (R_\mathrm{P,min})$,%
\footnote{In spite of the word ``variance'' used by Pendry \cite{pendry_jphysc_1980}, ``var'' is not a variance in the statistical sense (square deviation); it is a linear measure of uncertainty and should be better read as ``variation''.}
 with
\begin{equation}
  \mathrm{var} (R_\mathrm{P,min}) = R_\mathrm{P,min} \sqrt{\frac{8|V_\mathrm{0i}|}{\Delta E}}\ ,
  \label{eq:varR}
\end{equation}
where $\Delta E$ is the combined energy span of all independent $I(E)$ curves used in the analysis (also known as the size of the experimental data base used for the theory-experiment comparison) \cite{pendry_jphysc_1980}.%
\footnote{This procedure of estimating the error bars of a fit parameter is visualized in figure \ref{fig:parabolas} (section \ref{sec:noisy}).}

Given all the advantages of $R_\mathrm{P}$ listed above, one could consider it a perfect $R$ factor for LEED $I(E)$. The following section shows its weaknesses. Section \ref{sec:mod} presents an improved $R$ factor (named $R_\mathrm{S}$) that does not suffer from the deficiencies discussed in section \ref{sec:deficiencies}, without sacrificing the advantages of $R_\mathrm{P}$.

\section{\label{sec:deficiencies}Deficiencies of Pendry's \textit{R} factor}

\subsection{\label{sec:yrev}Dissimilar curves can have $R_\mathrm{P}=0$}
\begin{figure}
\includegraphics[width=7.5cm]{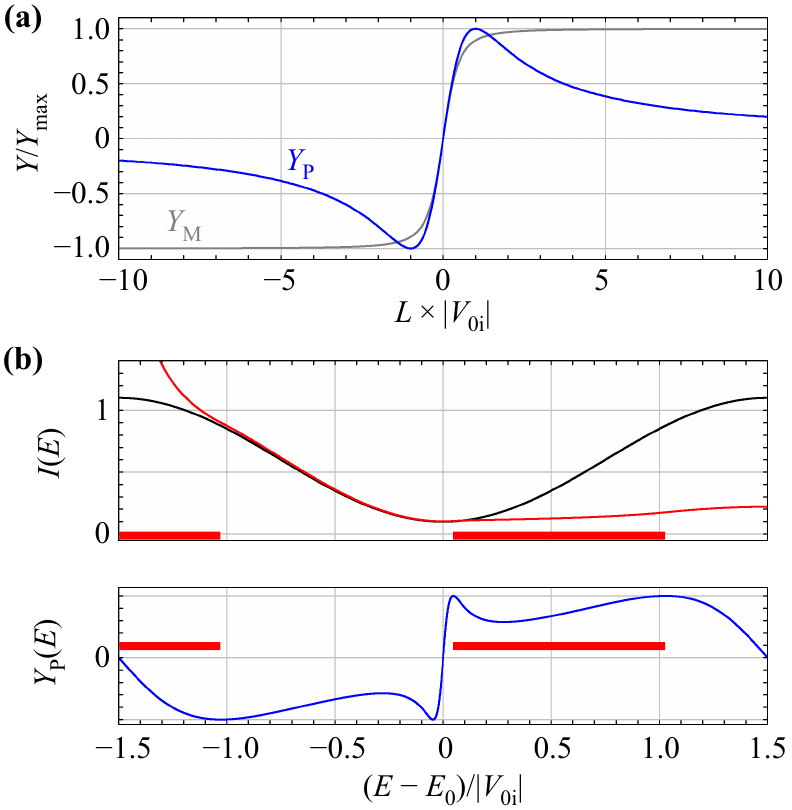}
\caption{\label{fig:YPe}(a) The $Y(L)$ function of Pendry's $R$ factor (blue). The gray curve shows an invertible function, $Y_\mathrm{M}$, which forms the basis for the $R$ factor $R_\mathrm{S}$ presented in this work. The horizontal axis is the scaled logarithmic derivative $L$ of the intensity [equation (\ref{eq:L})]. (b) The $I(E)$ plot shows two curves (black, red) that share the same $Y_\mathrm{P}(E)$ function, shown in the bottom. This implies that Pendry's $R$ factor between the two $I(E)$ curves is $R_\mathrm{P}=0$, a value that should occur only if two curves are identical except for a constant scale factor. The red bars indicate where the two $I(E)$ curves map onto different branches of the $Y_\mathrm{P}$ function; in these regions the two $I(E)$ curves are dissimilar (not proportional to each other). The horizontal axis is the scaled energy, centered at the position of a minimum of the $I(E)$ curves. For $|V_\mathrm{0i}|=4$\,eV, the plots in (b) would span 12\,eV on the energy axis.}
\end{figure}

As mentioned above, the $Y_\mathrm{P}$ function, equation (\ref{eq:YPe}), reaches a value of 0 for both, $L=0$ and $L\to \pm\infty$; its extrema of $\pm 1/|2V_\mathrm{0i}|$ are reached at $L=\pm 1/|V_\mathrm{0i}|$ (figure \ref{fig:YPe}a). This means that a given $Y_\mathrm{P}$ value can be reached for two different values of the logarithmic derivative $L$ (except for the two extrema of $Y_\mathrm{P}$). In other words, $Y_\mathrm{P}$ is not invertible. Therefore, one can obtain the same $Y_\mathrm{P}(E)$ curve with two different $I(E)$ curves, and Pendry's $R$ factor between these two curves will be $R_\mathrm{P} = 0$. This is exemplified in figure \ref{fig:YPe}b, where two qualitatively different synthetic $I(E)$ curves (black and red) share the same $Y$ function (blue).
%
The plot was constructed by assuming the black $I(E)$ curve (a raised cosine function), calculating the corresponding $L$ and $Y_\mathrm{P}$ functions, and then the inverse functions, taking the wrong branch of the inverse $Y_\mathrm{P}$ function in the regions marked by the red bars. This results in the red $I(E)$ curve. It is obvious that the black and red $I(E)$ functions are not proportional to each other and the $R$ factor between them should clearly be different from 0. Nevertheless, since both $I(E)$ functions share the same $Y_\mathrm{P}$ function, $R_\mathrm{P}=0$ for these two dissimilar functions. One may argue that it is unlikely that such a case will occur as a pure coincidence. However, it is easy to envisage a case where a region of the experimental $I(E)$ curve looks similar to one of the two $I(E)$ curves in figure \ref{fig:YPe}b, and the calculated one is similar to the other curve. Then, when minimizing the $R$ factor, the $I(E)$ curve in this energy region will ``pull'' the minimization process in a wrong direction, towards the correspondence given by the wrong branch of the $Y_\mathrm{P}$ function.

\subsection{\label{sec:minima}Pendry's $R$ factor is too sensitive to small intensity offsets}

Experimental $I(E)$ data are typically obtained by integrating the intensity of a diffraction maximum in a camera image of the LEED screen, after subtracting the background intensity caused by inelastically scattered electrons and stray light (e.g., from other bright spots) \cite{roucka_vac_2002,schmid_viperleed_2025}. The background intensity is not homogeneous and its exact distribution is not known, thus the background subtraction is inaccurate, especially for weak spots. In some cases, the intensity obtained this way can even become negative. Due to the uneven background, this problem also exist when fitting the peaks with a Gaussian (or similar function) \cite{mayer_compphyscomm_2012} since the curvature of the background intensity will lead to a nonzero fit amplitude even in the absence of any true peak. On the computational side, the intensity reaches zero when the complex amplitude of the outgoing wave (as a function of $E$) passes exactly through the origin of the complex plane. Even tiny changes of the complex amplitude that would not be noticeable in other places can make a qualitative difference, whether the complex amplitude passes through the origin or misses that point. Since the minimum intensity cannot be determined with sufficient accuracy, an $R$ factor should not be overly sensitive to a small offset of the intensity at a deep minimum.

Figure \ref{fig:atMin}a shows synthetic $I(E)$ curves (again, raised cosine functions) with different intensity offsets (thus different intensities at the minimum); the corresponding $Y_\mathrm{P}$ functions of Pendry's $R$ factor are shown in figure \ref{fig:atMin}b. If the intensity at the minimum is exactly zero, $Y_\mathrm{P}(E)$ is an almost perfectly straight line with a moderate slope in the region around the intensity minimum (black curve). When adding a very small intensity offset, the slope of $Y_\mathrm{P}(E)$ becomes very high, and extrema occur next to the energy of the intensity minimum. Thus, a small intensity offset (within the uncertainty of experimental $I(E)$ curves) leads to a large change of the $Y$ function (dark blue vs.\ black in figure \ref{fig:atMin}b). This will strongly affect the $R$ factor. As explained above, such an overly strong response to small intensity offsets is an undesirable property of an $R$ factor.

One should also note that an intensity minimum that reaches exactly zero has a well-defined position on the energy axis, while the position of a shallow minimum is usually less certain. Nevertheless, the black $Y_\mathrm{P}$ function in figure \ref{fig:atMin}b has a very gentle slope, which means that $R_\mathrm{P}$ is much less sensitive to the position of this minimum than to the positions of shallower minima, where the $Y_\mathrm{P}$ function is very steep (blue to orange). Near the minimum, a further problem becomes obvious in figure \ref{fig:atMin}b: The $Y_\mathrm{P}$ function of the minimum reaching zero intensity (black) is similar to that of the very shallow minimum shown in red. This is related to the fact that the $Y_\mathrm{P}(L)$ function shown in equation (\ref{eq:YPe}) and figure \ref{fig:YPe}a is non-invertible; this is a similar case as discussed in section \ref{sec:yrev}.

\subsection{\label{sec:noisy}Pendry's $R$ factor is noisy}

\begin{figure}[tb]
\includegraphics[width=8.5cm]{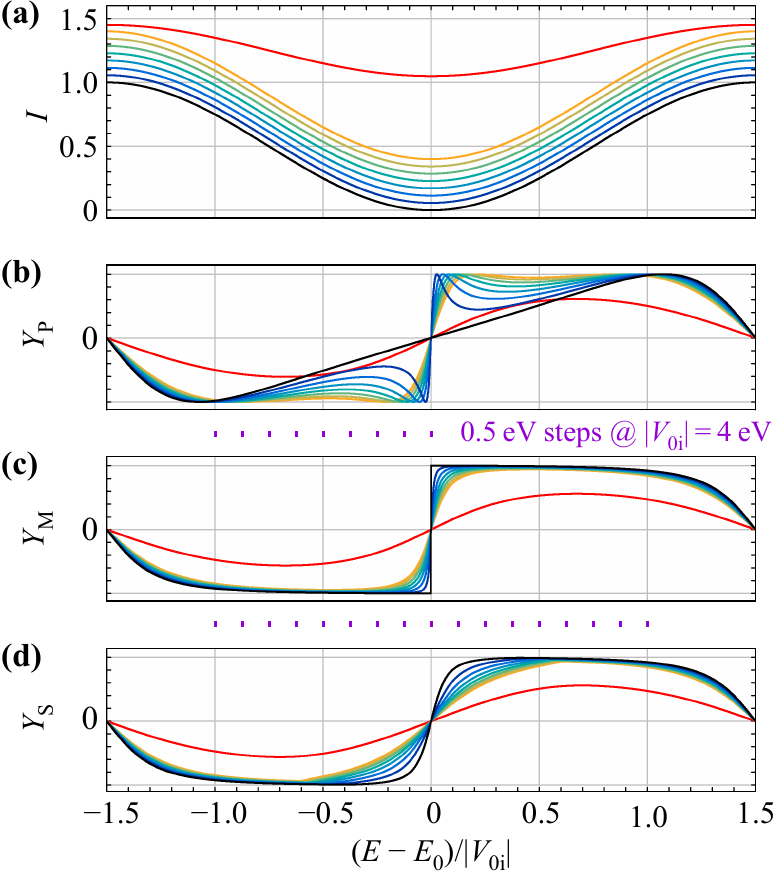}
\caption{\label{fig:atMin}Intensities and various $Y$ functions at a minimum with different intensity offsets. Note that all $Y$ functions are sensitive to relative changes only, but invariant to a scale factor of the intensity. Thus, the red intensity curve in (a) would be equivalent to a curve with a peak-valley height of 1.0 (like the other curves), but an intensity of 2.6 at the minimum. (b) For small, but non-zero intensity offset, $Y_\mathrm{P}$ of Pendry's $R$ factor reaches sharp minima and maxima next to the minimum (``cusps'').  The dots indicate the sampling interval in the case of 0.5\,eV energy steps and $|V_\mathrm{0i}|=4$\,eV. (c) The modified $Y_\mathrm{M}$ function of equation (\ref{eq:Ymod}) shows a hard step at an intensity minimum with a low or negligible offset (black curve), which is also undesirable. The $Y_\mathrm{S}$ function in (d) avoids these issues and depends smoothly on the intensity offset at the minimum.}
\end{figure}
%

\begin{figure}[tb]
\includegraphics[width=8cm]{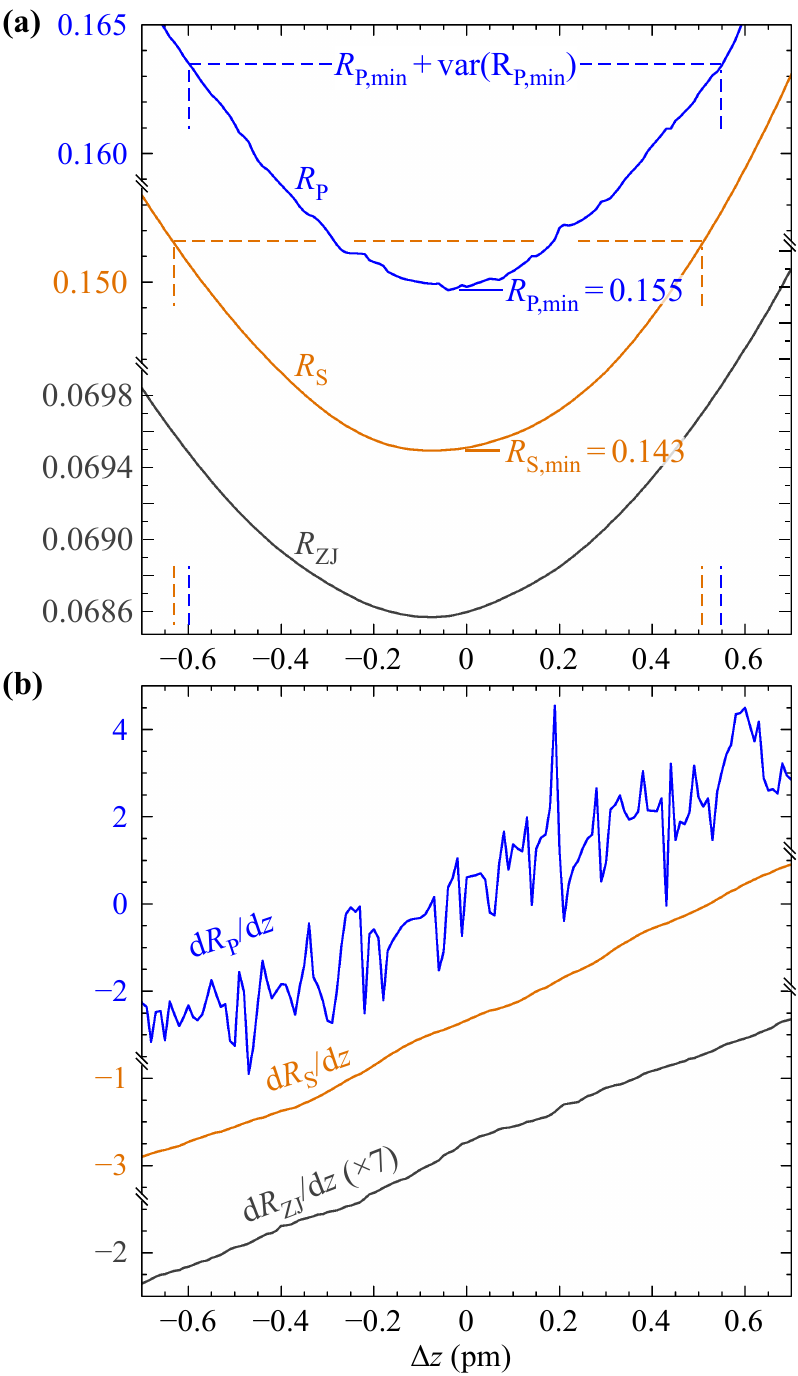}
\caption{\label{fig:parabolas}(a) $R$ factors and (b) their derivatives near the minimum. The horizontal axis is the displacement $\Delta z$ of the two symmetry-equivalent uppermost Fe atoms of the  $\alpha$-Fe$_2$O$_3(1{\bar{1}02)}$-$(1\times 1)$ structure \cite{kraushofer_viperleed_2025}. The vertical axis for the Zanazzi--Jona $R$ factor has been scaled by a factor of 7 since the curvature of $R_\mathrm{ZJ}$ vs.\ $\Delta z$ is lower than that of the two other $R$ factors by roughly this factor. The dashed lines indicate the error bar of the parameter $\Delta z$ derived from $\mathrm{var}(R)$, equation (\ref{eq:varR}).
The plot does not include $R_2$; it is smooth, but its minimum would lie close to the right border of the plot. This is consistent with a previous study finding less accurate results when minimizing $R_2$ \cite{sporn_accuracy_1998}.
 }
\end{figure}

For structure optimization, the $R$ factor is the objective function that should be minimized by varying the parameters of the structural model of the surface. Many minimization algorithms are sensitive to noise; this is especially true for gradient-based methods, which offer the best performance as soon as a rough (exploratory) search has identified the region of the global minimum \cite{imre_jax_2026}. It has been noted early that minimizing $R_\mathrm{P}$ is plagued by noise \cite{rous_prss_1992}. The noisiness of $R_\mathrm{P}$ as a function of structural parameters was initially attribute to numerical noise in the calculation of the $I(E)$ curves \cite{rous_prss_1992}, but it is easy to show that noise in $R_\mathrm{P}$ is present also when numerical noise of the underlying $I(E)$ calculations can be excluded.
Figure \ref{fig:parabolas}a shows Pendry's $R$ factor of a theory--experiment comparison for  $\alpha$-Fe$_2$O$_3(1{\bar{1}}02)$-$(1 \times 1)$ \cite{kraushofer_viperleed_2025}. $R_\mathrm{P}$ is plotted as a function of a single atomic coordinate,  the $z$ position of the topmost two (symmetry-equivalent) Fe atoms. The calculated $I(E)$ curves were not obtained by full-dynamic or tensor LEED  \cite{rous_prss_1992} calculations (which might lead to numerical noise) but by a linear interpolation of the complex amplitudes between $\Delta z = -1\,\mathrm{pm}$ and $\Delta z = +1\,\mathrm{pm}$.%
\footnote{Such a linear approximation is justified for very small geometric displacements of a few picometers \cite{imre_jax_2026}. For larger displacements, a tensor LEED calculation \cite{rous_prss_1992} or a full-dynamic LEED calculation would be required for each displacement.}
We took this approach because a simple linear interpolation cannot lead to numerical noise of the $I(E)$ curves as a function of the parameter. Nevertheless, figure \ref{fig:parabolas} shows that $R_\mathrm{P}$ is noisy; it does not exhibit a smooth, parabola-like minimum. 
The roughness of Pendry's $R$ factor becomes even more obvious when looking at the gradient; it is clear that optimization algorithms using this gradient will struggle finding the minimum  (figure \ref{fig:parabolas}b; the gradient was evaluated numerically with 0.01\,pm spacing of the points).

\begin{figure}
\includegraphics[width=8.5cm]{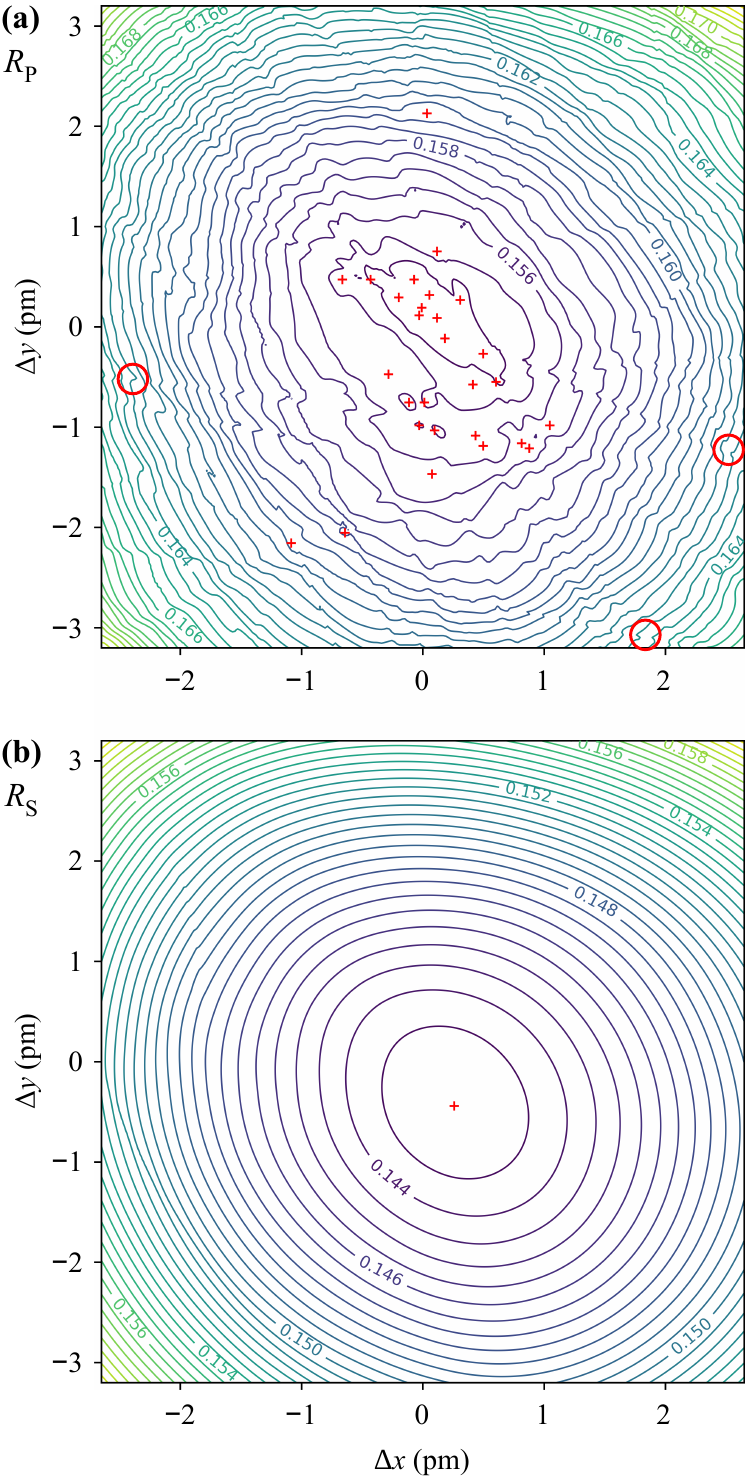}
\caption{\label{fig:contours}Contour plots of (a) $R_\mathrm{P}$ and (b) $R_\mathrm{S}$ as a function of the $x$ and $y$ displacement of the upper two symmetry-equivalent Fe atoms of $\alpha$-Fe$_2$O$_3(1{\bar{1}02)}$-$(1\times 1)$, based on tensor-LEED calculations. The red circles mark a few points where the gradients of $R_\mathrm{P}$ are grossly misleading, even far from the minimum. Minima are marked by small, red crosses. }
\end{figure}

One might argue that the parameter range of $\pm 0.7$\,pm in figure \ref{fig:parabolas}  is very small. This range is only slightly larger than the uncertainty of the parameter $\Delta z$ determined via $\mathrm{var} (R_\mathrm{P,min})$ [cf.\ equation (\ref{eq:varR})]. The gradient near the boundaries of this range points in the right direction, so one should find the minimum in spite of the noise. This may be true for a one-dimensional optimization problem, but LEED $I(E)$ studies require high-dimensional optimization, typically with dozens, sometimes even more than 100 parameters \cite{kisslinger_prb_2023}. To illustrate the impact of the noise of $R_\mathrm{P}$ on multidimensional optimization, figure \ref{fig:contours}a shows a contour plot of $R_\mathrm{P}$ as a function of two optimization parameters (the $x$ and $y$ coordinates of an atom). The plot shows that the gradients of $R_\mathrm{P}$ can be grossly misleading also far from the minimum (red circles). High-dimensional minimization is much more demanding than finding a minimum in one or two dimensions \cite{chen_dimensionality_2015}. Many minimization algorithms use the difference of the gradients between successive iteration points to construct an approximation for the Hessian of the minimization problem \cite{nocedal_book_2006}. If the gradients are strongly affected by noise, this approximation for the Hessian will be grossly wrong. Then, in addition to getting trapped in local minima (like those marked by red crosses in figure \ref{fig:contours}a), minimization algorithms may get stuck at off-minimum parameter values if the algorithm cannot cope with a noisy objective function with seemingly inconsistent gradients.

The noise of $R_\mathrm{P}$ is caused by the extrema of $Y_\mathrm{P}$ near a minimum (figure \ref{fig:atMin}b). As mentioned in section \ref{sec:minima}, the occurrence of these extrema depends sensitively on whether the intensity at the minimum reaches exactly zero or not. In practice, another important issue is the fact that the integration of equation (\ref{eq:RPe}) in current LEED codes is performed as a discrete sum, with an energy step of typically 0.5 or 1\,eV.%
\footnote{The energy step is usually that of the experimental $I(E)$ curves, i.e. the energy step between successive images of the LEED screen. The calculations are performed with larger steps and the calculated $I(E)$ curves are interpolated to the experimental step size. The ViPErLEED Spot tracker package \cite{schmid_viperleed_2025} contains a utility for interpolation to smaller steps, which may be used for the experimental data, if desired. It has been noted early that a fine grid like 0.25\,eV can reduce the noisiness of $R_\mathrm{P}$ \cite{van_hove_ssr_1993}.}
Figure \ref{fig:atMin}b shows that even 0.5\,eV steps are rather coarse compared to the sharpness of the sharpest extrema (the cusps) of $Y_\mathrm{P}$ (dark blue curve). Thus, depending on the position of the sampling points with respect to the cusps, a slight energy shift can lead to large changes of the $Y_\mathrm{P}$ values at the points sampled. This is another source of noise, which comes in addition to the overly sensitive response to the residual intensity at deep minima.

\section{\label{sec:mod}Modifying Pendry's $R$ factor}

The results of section \ref{sec:deficiencies} point at the $Y_\mathrm{P}$ function as the root of most deficiencies of Pendry's $R$ factor, especially the fact that $Y_\mathrm{P}$  ``folds back'' towards 0 after reaching its extrema, which makes it non-invertible: the inverse function is ambiguous. It is easy to create modified $Y$ functions that do not fold back for high values of $|L|$ and still suppress the infinity of $L$ at $I \to 0$; e.g.
\begin{equation}
  Y_\mathrm{M} = \frac{L}{\sqrt{1+4V_\mathrm{0i}^2L^2}} = \frac{I'}{\sqrt{I^2+4V_\mathrm{0i}^2I'^2}}
  \label{eq:Ymod}
\end{equation}
This function is shown as the gray curve in figure \ref{fig:YPe}a.
$Y_\mathrm{M}(L)$ and the other $Y$ functions discussed below have the same slope as $Y_\mathrm{P}(L)$ near $L=0$, and also the minimum and maximum function values are the same for all $Y$ functions, $\pm 1/|2V_\mathrm{0i}|$.

$Y_\mathrm{M}(L)$ avoids the sharp cusps near a deep intensity minimum and the overly strong dependence on small intensity offsets, see figure \ref{fig:atMin}c. This figure shows an undesirable property of $Y_\mathrm{M}$, however: For minima with a small or vanishing offset (black and dark blue curves), $Y_\mathrm{M}$ has a step at the position of the minimum. When using the $R$ factor to compare two curves, the position of the minimum will usually be slightly different between the two curves. In the energy range between the minima of the two curves, their $Y_\mathrm{M}$ functions will be almost maximally different, $|Y_\mathrm{exp}-Y_\mathrm{th}|\approx 2Y_\mathrm{max}$. If the energy difference of the minima is less than the energy step used in the numerical evaluation, it comes down to coincidence whether one of the discrete energies used in the evaluation is between the minima of the two curves. If so, this energy will lead to a clear increase of the $R$ factor. Otherwise, if the energy grid happens to have no point the region between the minima of the two curves, the difference will affect the $R$ factor only marginally. As an example, when using 0.5\,eV energy steps, an 0.4\,eV difference between the minima of the two curves may go unnoticed in some cases, while in other cases an 0.1\,eV difference gets penalized. It is clear that such an ``all or nothing'' measure with some degree of arbitrariness  is not suitable as an objective function for minimization.

In principle, the steps of $Y_\mathrm{M}$ at deep minima of $I$ could be avoided by smoothing the $Y_\mathrm{M}$ function. This approach, however, would introduce a problem at the beginnings and ends of an $I(E)$ curve: smoothing would require to access data outside the available range. Linear extrapolation (with weights quickly decreasing with increasing distance from the edge) followed by smoothing can sometimes help \cite{schmid_sg_2022}, but is not applicable in the current case.%
\footnote{Correct extrapolation of $Y_\mathrm{M}$ is not feasible: Assume a deep intensity minimum barely outside the boundary of the data (the end of the energy range). Next to the boundary, $Y_\mathrm{M}$ would be very close to its extremum, $+Y_\mathrm{max}$ or $-Y_\mathrm{max}$; its sharp transition to the other sign would be at the position of the minimum, slightly outside of the boundary. The smoothed $Y_\mathrm{M}$ function should tend towards zero already inside the boundary, to reach exactly zero at the position of the minimum. However, the $Y_\mathrm{M}$ function near the boundary of the data provides no indication of the presence of an intensity minimum with the sign jump of $Y_\mathrm{M}$. The only way out would be extrapolation of the intensities $I$ and calculating the out-of-bounds values of $Y_\mathrm{M}$ from the extrapolated $I(E)$. Linear extrapolation is not suitable near a minimum of $I(E)$; extrapolation with higher-order polynomials (at least 2nd order) would be required. Since the uncertainty of extrapolation increases with the number of free parameters and polynomials are prone to overshoot (even in interpolation \cite{dahlquist_numerical_2008}), polynomial extrapolation is not a robust method.}
Furthermore, since the $Y_\mathrm{M}$ functions for $I(E)$ curves with some offset are very similar to what one would obtain by smoothing the step-like $Y_\mathrm{M}$ obtained for zero offset (see figure \ref{fig:atMin}c), smoothing will make all the $Y$ functions very similar, irrespective of the offset. Such an almost complete independence on the offset is undesirable, since the offset carries information (on how close the complex amplitudes come to the origin of the complex plane). This information should not be discarded (which would be the other extreme to $Y_\mathrm{P}$'s overreaction on a small offset).

As an alternative to smoothing, a purely local approach would be preferable. One can avoid the division by (almost) zero at the minima by adding a term to the denominator. Since the $Y$ function of a Pendry-style $R$ factor must be scale-invariant with respect to multiplying the intensity by a constant factor, the only nonzero quantity locally available at a minimum and suitable for this purpose is the second derivative of the intensity $I''$. One could therefore consider the following $Y$ function:
\begin{equation}
  Y_\mathrm{M2} = \begin{cases}
      \frac{I'}{\sqrt{I^2+4V_\mathrm{0i}^2 I'^2+\eta V_\mathrm{0i}^4 I''^2}}&\text{if }I''> 0\\
      \frac{I'}{\sqrt{I^2+4V_\mathrm{0i}^2 I'^2}}&\text{otherwise.}
      \end{cases}
  \label{eq:Ymod}
\end{equation}
The constant $\eta$ determines by how much the steps of the $Y_\mathrm{M}$ functions are smoothed out. In practice, it turns out that $Y_\mathrm{M2}$ depends too weakly on the intensity offset, a problem that was already discussed above in the context of smoothing the $Y_\mathrm{M}$ function.

Before introducing an improved $Y$ function, let us discuss the desirable behavior of a $Y$ function near an intensity minimum. As mentioned above, it should avoid the step of $Y_\mathrm{M}$ (seen in figure \ref{fig:atMin}c). Since numerical evaluation is performed by sampling at discrete energy points, $Y$ should be roughly linear between two evaluation points.  Furthermore, the Y function should be sufficiently, but not excessively sensitive to the offset, i.e.\ to the intensity at the minimum. For a shallow minimum (large offset), the logarithmic derivative $L$ decreases in magnitude, and the slope of the $Y$ function should also decrease. This desired behavior cannot be accomplished with $L$ alone, because $L$ becomes very large towards the minimum.

Our approach makes use of the fact that the offset of a parabolic minimum can be expressed as $I_\mathrm{min} = I - \frac{I'^2}{2I''}$, and $I_\mathrm{min}/(V_\mathrm{0i}^2 I'')$ provides a dimensionless measure for the offset at a minimum (for how shallow a minimum is). We use this expression to obtain the desired behavior of the $Y_\mathrm{S}$ function:
\begin{align}
  y_1 &= \frac{\alpha}{V_\mathrm{0i}^2}\left(\frac{I}{I''} - \frac{I'^2}{2I''^2}\right) + \beta \label{eq:y1}\\
  &(\text{with } \alpha=4, \beta=0.15) \label{eq:alphabeta}\\
  y_2 &= \frac{y_1}{\sqrt{1 + y_1^2}}  \label{eq:y2}\\
  Y_\mathrm{S} &= \begin{cases}
      \frac{I'}{\sqrt{I^2+4V_\mathrm{0i}^2 I'^2+ y_2^2 V_\mathrm{0i}^4 I''^2}}&\text{if }I''> 0\text{ and }y_1 > 0\\
      \frac{I'}{\sqrt{I^2+4V_\mathrm{0i}^2 I'^2}}&\text{otherwise.}
      \end{cases} 
    \label{eq:Ys}
\end{align}
Similar to $\eta$ in equation (\ref{eq:Ymod}), $y_2$ determines how much the step of $Y_\mathrm{M}$ is smoothed out, or, in other words, how much the slope of the $Y$ function at a minimum gets reduced. 
As described above, the first term in equation (\ref{eq:y1}) is a scale-invariant measure of the offset; thus the smoothing (slope-limiting) action of equation (\ref{eq:Ys}) increases with increasing offset (shallower minima). Equation (\ref{eq:y2}) limits this smoothing for very shallow minima because these yield only a weak modulation of the $Y$ function without any additional measures (see the red curves in figure \ref{fig:atMin}; there, the logarithmic derivative $L=I/I'$ is sufficiently small). The constant $\beta$ in equation (\ref{eq:y1}) provides a minimum degree of smoothing (maximum slope) of the $Y$ function at deep minima where the intensity approaches or reaches zero. The case distinction in equation (\ref{eq:Ys}) ensures that the correction is restricted to regions close to the minima; checking for the sign of $y_1$ (or, equivalently, $y_2$) provides a smooth transition to the unmodified $Y_\mathrm{M}$ function ($y_1=y_2=0$ at the transition).

Equation (\ref{eq:y1}) has two parameters: $\alpha$, which determines the influence of the offset of the minimum, and $\beta$, which determines the behavior at a minimum reaching $I=0$. The choice of $\alpha=4$ and $\beta=0.15$ is motivated by the desired behavoir of an $R$ factor discussed above. With these values, as figure \ref{fig:atMin}d shows, (i) there are clear differences between the curves with zero offset (black) and larger offsets (blue to orange), and (ii) the steepest curve (black) at the intensity minimum varies sufficiently slowly to yield a smooth variation of the $R$ factor when sampled at discrete energy with 0.5\,eV steps (violet tick marks in figure \ref{fig:atMin}). Since the minimum shown in figure \ref{fig:atMin} is roughly the sharpest that can occur in real $I(E)$ data and a low value of $|V_\mathrm{0i}|= 4$\,eV was assumed, the figure should be seen as a worst case; for higher $|V_\mathrm{0i}|$ values, the tick spacing would be shorter and for wider minima the $Y_\mathrm{S}$ function gets smoothed out. Nevertheless, we recommend using an energy step of 0.5\,eV or less also for larger values of $|V_\mathrm{0i}|$.

Apart from the behavior near the minima, the $Y_\mathrm{S}$ function is very similar to $Y_\mathrm{P}$, thus one can use it in equation (\ref{eq:RPe}). We call the $R$ factor obtained this way $R_\mathrm{S}$. The subscripts ``S'' stands for ``smooth'', which is justified since $R_\mathrm{S}$ is not plagued by the noise of $R_\mathrm{P}$, see figures \ref{fig:parabolas} and \ref{fig:contours}. Both $R_\mathrm{S}$ and it's derivative react smoothly to small changes of structure parameters.
\section{\label{sec:compare}Comparison of $R$ factors}

\subsection{\label{sec:similar}Similarities of $R_\mathrm{S}$ and $R_\mathrm{P}$}

\begin{figure*}[tb]
\includegraphics[width=\textwidth]{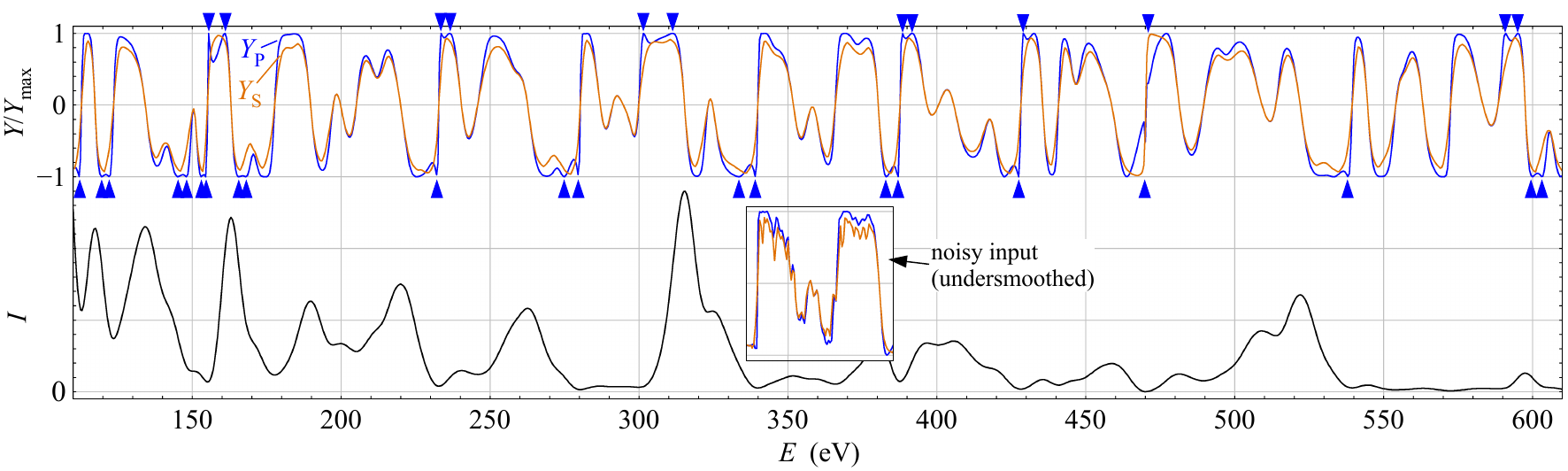}
\caption{\label{fig:IEcurve}Experimental $I(E)$ curve and its $Y$ functions. The black curve at the bottom is $I(E)$ of the (1,2) beam of $\alpha$-Fe$_2$O$_3(1{\bar{1}02)}$-$(1\times 1)$. The upper curves are the $Y$ functions of Pendry's $R$ factor $R_\mathrm{P}$ (blue) and of the $R$ factor $R_\mathrm{S}$ introduced in this work (orange). Blue arrows mark cusps of $Y_\mathrm{P}$, where the $Y_\mathrm{P}(L)$ function reaches an extremum and folds back (cf.\ figure \ref{fig:YPe}a). The inset around $E=350$\,eV shows the $Y$ functions affected by experimental noise due to insufficiently smoothed data (smoothing parameter 1.0\,eV, far below the recommended range \cite{schmid_viperleed_2025}).}
\end{figure*}

\begin{figure}[tb]
\includegraphics[width=7.5cm]{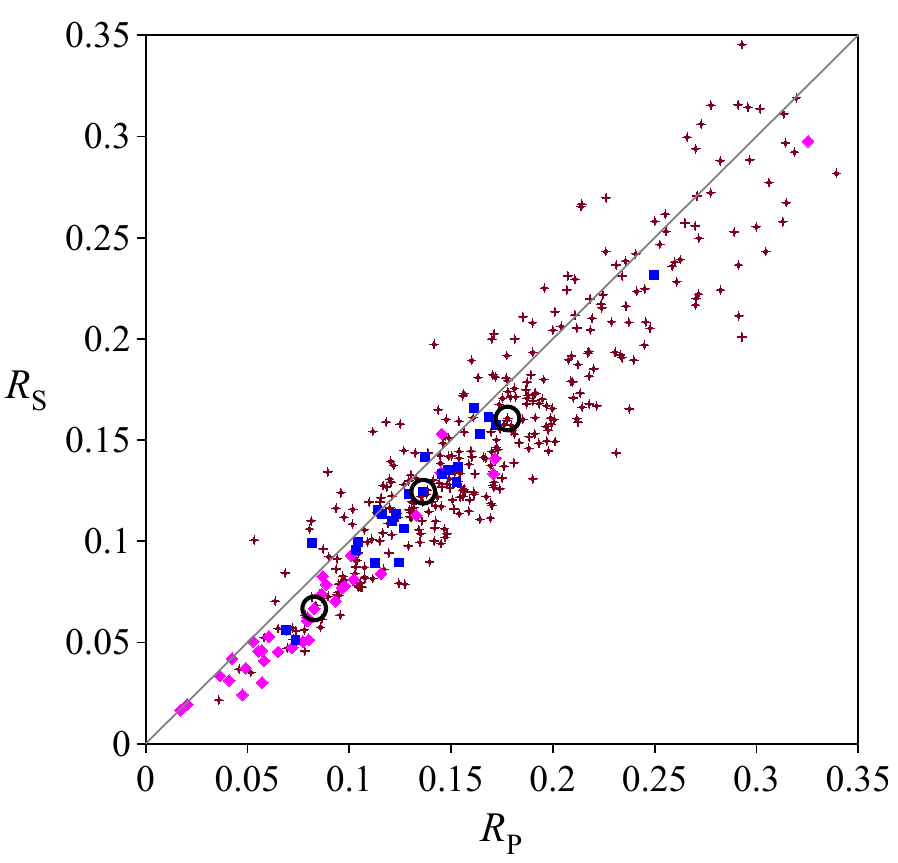}
\caption{\label{fig:scatter}Correlation between $R_\mathrm{S}$ and $R_\mathrm{P}$. Each data point gives the $R$ factors between the calculated and experimental $I(E)$ curve of a particular beam; also the overall $R$ factors for all beams are included (marked by black circles). The data are from three systems, $\alpha$-Fe$_2$O$_3(1{\bar{1}02)}$-$(1\times 1)$ \cite{kraushofer_viperleed_2025} (blue squares), Ir(100)-$p(3 \times 1)$-MnO$_2$ \cite{ferstl_prl_2016} (pink diamonds), and  Pt(111)-$(10 \times 10)$-49Te \cite{kisslinger_prb_2023} (brown crosses).}
\end{figure}

Except for the minima of the intensity, the $Y$ functions of Pendry's $R$ factor  $R_\mathrm{P}$ and the newly presented  $R_\mathrm{S}$ are very similar, as shown in figure \ref{fig:IEcurve}, which presents these two $Y$ functions for a typical experimental $I(E)$ curve. Therefore, it is not astonishing that the values of these two $R$ factors are also similar. As figure \ref{fig:scatter} shows, on average, $R_\mathrm{S}$ values are slightly lower than $R_\mathrm{P}$, because $R_\mathrm{S}$ is not plagued by the cusps at deep minima, which exaggerate minuscule differences of the $I(E)$ curves.

Since the new $R$ factor $R_\mathrm{S}$ is based on the same construction principle as $R_\mathrm{P}$ and has similar values, one can also determine the error bars for fit parameters the same way as for $R_\mathrm{P}$, using $\mathrm{var}(R_\mathrm{min})$ as defined in equation (\ref{eq:varR}), by simply replacing $R_\mathrm{P}$ with $R_\mathrm{S}$ in this equation. This is justified since the derivation of equation (\ref{eq:varR}) in Ref.\ \cite{pendry_jphysc_1980} is based on peaks only, where the $Y_\mathrm{M}$ and $Y_\mathrm{S}$ functions are essentially identical (the regions with negative slope of the $Y$ functions in figure \ref{fig:IEcurve}). Since the numerical $R_\mathrm{P}$ and $R_\mathrm{S}$ values are similar (figure \ref{fig:scatter}) and the curvatures of the $R$ factor at the parameter minima are also very similar, using the procedure based on equation (\ref{eq:varR}) results in very similar error estimates, see figure \ref{fig:parabolas}.

\subsection{\label{sec:noise}Sensitivity to experimental noise}

In contrast to $Y_\mathrm{P}$, the $Y_\mathrm{S}$ function depends on the second derivative $I'' = \mathrm{d}^2I/\mathrm{d}E^2$, though only in the vicinity of sufficiently deep minima. Since most sources of noise in the experimental $I(E)$ curves increase with increasing intensity (e.g., shot noise, modulation by the grids), this dependence of $Y_\mathrm{S}$ on $I''$ is less severe than for $R_\mathrm{ZJ}$, which is sensitive to $I''$ also at the maxima. This means that the noise of the second derivative $I''$ will have a smaller effect on $R_\mathrm{S}$ than on $R_\mathrm{ZJ}$. In practice, the $Y_\mathrm{S}$ functions of moderately noisy (weakly smoothed) $I(E)$ curves do not show any noticeable increase of the noise at deep minima where $I''$ comes into play.

Concerning noise, the main difference between  $R_\mathrm{S}$ and $R_\mathrm{P}$ is not the use of the second derivative. As figure \ref{fig:YPe}a shows, the $Y_\mathrm{P}(L)$ function reaches saturation earlier than $Y_\mathrm{M}(L)$. ($Y_\mathrm{S}$ and $Y_\mathrm{M}$ are identical with the exception of the regions around deep minima.) When $Y(L)$ reaches saturation, i.e.\ when the derivative of $\mathrm{d}Y/\mathrm{d}L$ vanishes, the noise of the underlying logarithmic derivative $L$ is suppressed. It turns out that $L$ values near $1/|V_\mathrm{0i}|$, where $Y_\mathrm{P}(L)$ reaches saturation, are very common. In figure \ref{fig:IEcurve}, these are the regions where $Y_\mathrm{P}$ reaches an extremum of $\pm Y_\mathrm{max}$; this occurs next to all intensity minima except for the very shallow ones. In these regions, $Y_\mathrm{S}$ does not yet reach its saturation value $\pm Y_\mathrm{max}$, which means that the slope of $Y_\mathrm{S}(L)$ is nonzero and any noise of the $I(E)$ data (and the corresponding logarithmic derivative $L$) will affect $Y_\mathrm{S}$ to some degree. This can be seen in the inset of figure \ref{fig:IEcurve}, where the $Y$ functions from noisy (insufficiently smoothed) experimental data are plotted. Nevertheless, in practice, the impact of this higher noise sensitivity of $Y_\mathrm{S}$ is marginal. First, it is partly compensated by regions where $Y_\mathrm{P}$ is more sensitive to noise, since the cusps at deep intensity minima are sensitive to small changes of the input data, as discussed in section \ref{sec:minima}. Second, the noise of $Y$ in the near-saturation regions is less than in regions where the logarithmic derivative is small and the slope of $Y(L)$ is much higher. Since the initial slopes of $Y_\mathrm{S}(L)$ and $Y_\mathrm{P}(L)$ are identical, there is no difference between the two $R$ factors in these regions, which often dominate the noise.
 The comparison of the $R$ factors between experimental and calculated $I(E)$ curves in figure \ref{fig:scatter} also indicates that $R_\mathrm{S}$ is not plagued by experimental noise: Since the $Y$ functions of both $R$ factors are very similar, a higher impact of noise in typical experimental data on $R_\mathrm{S}$ would lead to higher $R_\mathrm{S}$ values compared with $R_\mathrm{P}$. In reality, the opposite is true: on average, $R_\mathrm{S}$ is lower than $R_\mathrm{P}$. Only for very noisy data, obtained by far too weak smoothing,%
 \footnote{The recommended smoothing parameter range in the I(V) Curve Editor of the ViPErLEED package \cite{schmid_viperleed_2025} is $0.6V_\mathrm{0i}$ to $1.3V_\mathrm{0i}$. Smaller values of the smoothing parameter do not affect the signal but only increase the noise. For realistic experimental data, noise causes $R_\mathrm{S}$ values higher than $R_\mathrm{P}$ only for smoothing parameters well below $0.5V_\mathrm{0i}$. The smoothing parameter is roughly inversely proportional to the band width of the filtering operation \cite{schmid_viperleed_2025,schmid_sg_2022}.}
the situation is reversed, with $R_\mathrm{S}$ higher than $R_\mathrm{P}$.

The ``early'' saturation of $Y_\mathrm{P}$ reached in many regions of the $I(E)$ curves may suppress noise to a larger degree than the weak, but nonzero slope of $Y_\mathrm{M}(L)$ [or equivalently $Y_\mathrm{S}(L)$], at the same $L$ values. On the other hand, this vanishing derivative is a disadvantage of Pendry's $R$ factor. It means that $R_\mathrm{P}$ is insensitive to the slope of the $I(E)$ curves in these regions. In other words, information present in the slopes of $I(E)$ is discarded by Pendry's $R$ factor. The $Y_\mathrm{S}$ function reaches saturation only at higher absolute values of the logarithmic derivative $L$; these values are reached only close to deep minima. There, the construction of the $Y_\mathrm{S}$ function given by equation (\ref{eq:y1})--(\ref{eq:Ys}) ensures that $Y_\mathrm{S}$ contains information on the depth of the minima, as discussed in section \ref{sec:mod} and exemplified in figure \ref{fig:atMin}d. Figure \ref{fig:IEcurve} shows that $Y_\mathrm{S}$ for realistic data reaches its saturation value $\pm Y_\mathrm{max}$ at very few points, which means that the information lost at these few points is negligible.

\subsection{\label{sec:accuracy}Which $R$ factor yields the  most accurate result?}

The new $R$ factor, $R_\mathrm{S}$, may be slightly more sensitive than $R_\mathrm{P}$ to strong noise in the experimental data. The relevant question, however, is not the influence of noise on the $R$ factor but to which degree the outcome of the structure optimization is affected by noise and other imperfections of the experimental data. The $R$ factor should not steer the minimization away from the true value in the presence of noise and imperfections. Figures \ref{fig:parabolas} and \ref{fig:contours} show that minimization of different $R$ factors leads to slightly different  parameter values.
Unfortunately, we cannot determine which of these results is closer to reality, since there is no method to determine the structure with better accuracy than a LEED $I(E)$ study. We therefore take a different approach: We use data with more imperfections than our standard data set for $\alpha$-Fe$_2$O$_3(1{\bar{1}02)}$-$(1\times 1)$ and examine how well the resulting fit  parameters agree with the result of our best data set. Minimizing a given $R$ factor with imperfect data should yield fit parameters that are as close as possible to the result obtained with the best data.

For this comparison, 30 structural parameters (all independent atom coordinates in the upper two layers) of $\alpha$-Fe$_2$O$_3(1{\bar{1}02)}$-$(1\times 1)$ have been optimized with different experimental $I(E)$ data. We analyze the resulting deviations of the parameters between the optimization with the imperfect data and the best data set for a given $R$ factor. Since different parameters influence the $R$ factor to a different degree,%
\footnote{Weak scatterers and atoms deeply buried have a weak influence on the $R$ factor. Also, displacements in $x$ and $y$ direction have a weaker influence than vertical displacements.}
the deviations have been normalized: Each deviation has been divided by the uncertainty of the respective parameter, derived from $\mathrm{var} (R_\mathrm{P,min})$, cf.\ equation (\ref{eq:varR}). Statistics of these 30 normalized parameter deviations, obtained by minimizing the three $R$ factors $R_\mathrm{P}$, $R_\mathrm{S}$, and $R_\mathrm{ZJ}$
have been plotted in figure \ref{fig:paramerrors}.%
\footnote{We do not include $R_2$ [equation (\ref{eq:R2})] since the resulting parameter values strongly deviate from those of the other $R$ factors and previous work shows that minimizing $R_2$ leads to less accurate results than those obtained with $R_\mathrm{P}$ \cite{sporn_accuracy_1998}.}
The results obtained with different imperfect $I(E)$ input data are discussed in the following paragraphs:

\begin{figure}[tb]
\includegraphics[width=8.5cm]{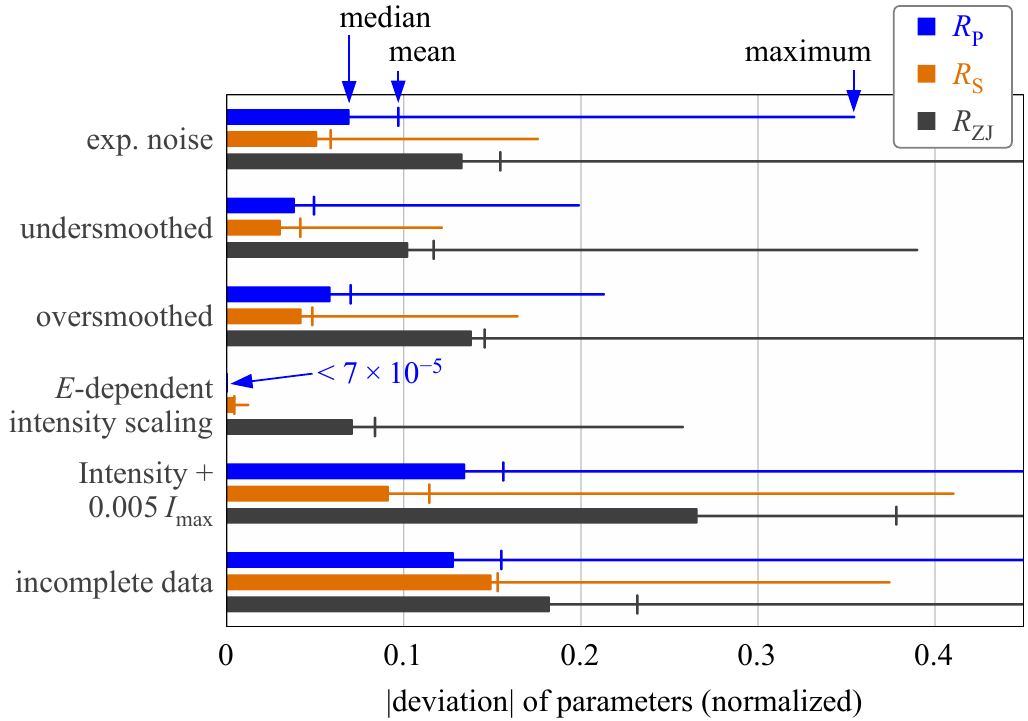}
\caption{\label{fig:paramerrors}Deviations of the best-fit parameters for data sets with imperfect $I(E)$ curves. The plot shows statistics of the normalized deviations of 30 fit parameters obtained using imperfect experimental data sets, with respect to the parameters derived from the best data set, when using a given $R$ factor for both. It therefore provides a measure for how sensitive structure optimization with a given $R$ factor is with respect to experimental imperfections. The horizontal axis is the absolute value of the deviation, normalized by the error bar for the respective parameter as derived from $\mathrm{var}(R_\mathrm{P})$. The plot shows the median, mean, and maximum of these normalized, absolute deviations. See the text for a description of the data sets named at the left.}
\end{figure}

\textit{Experimental noise}---The best data set (used as the reference) suppresses the impact of instrumental imperfections (modulation by the grids of the LEED optics, screen defects) and sensor noise by averaging over two measurements with the beams impinging on slightly different positions of the LEED screen. In addition, this reference data set was obtained using standard image processing techniques to correct for inhomogeneities of the LEED screen and sensor (flat field and dark frame corrections \cite{schmid_viperleed_2025}). The imperfect experimental $I(E)$ curves used for the top section of figure \ref{fig:paramerrors} were created without these corrections and without averaging, which results in increased noise. Figure \ref{fig:paramerrors}  shows that the parameters derived from these noisy $I(E)$ curves  show the least deviation from the reference when $R_\mathrm{S}$ is used for the optimization, closely followed by $R_\mathrm{P}$. $R_\mathrm{ZJ}$ performs worst. The difference between $R_\mathrm{P}$ and $R_\mathrm{S}$ is not necessarily significant since the values obtained with $R_\mathrm{P}$ might not exactly correspond to the global minimum, in spite of running all $R_\mathrm{P}$ minimizations with more than 50 different starting values (random starting points as well as the best-fit parameters of the reference set) and selecting the parameters corresponding to the best $R$ factor obtained in these $>50$ runs. (This problem is due to the noise of $R_\mathrm{P}$, see figure \ref{fig:contours}a and section \ref{sec:performance}.)

\textit{Undersmoothed $I(E)$ data}---The experimental data were based on the same measurement as the best data set, but the smoothing applied was much weaker than recommended \cite{schmid_viperleed_2025}. The plot summarizes the results obtained with two data sets, with smoothing parameters of 1.0 and 1.5\,eV (the recommended value is $>2.2$\,eV \cite{schmid_viperleed_2025}). Insufficient smoothing increases especially the high-frequency noise, which strongly affects the second derivative $I''$ \cite{tear_comparison_1981}. Since $R_\mathrm{P}$  is the only $R$ factor under consideration that does not make use of $I''$, one should expect that it is the least sensitive to this type of noise. Nevertheless, the parameter values obtained with $R_\mathrm{S}$ come closer to the reference value of the best data set than those from $R_\mathrm{P}$ minimization. Again, due to the noisiness of $R_\mathrm{P}$, the difference between $R_\mathrm{S}$ and  $R_\mathrm{P}$ is not necessarily significant.  $R_\mathrm{ZJ}$ leads to the largest deviations from the reference.

\textit{Oversmoothed $I(E)$ data}---The best data set was strongly smoothed by an additional smoothing step (smoothing parameter 6\,eV; this is more than recommended). This smoothing step eliminates fine detail, e.g., double peaks with a separation of ${\approx}2V_\mathrm{0i}$ turn into a single maximum. The sequence of the parameter deviations obtained with the different $R$ factors is the same as in the previously discussed cases, with $R_\mathrm{S}$ best and $R_\mathrm{ZJ}$ worst.

\textit{Energy-dependent intensity scaling}---This problem may occur in experiment due to improper (or missing) correction of the intensity data for the incident electron beam current $I_0$, or in the case of an energy-dependent gain of a channelplate LEED system. To simulate it, the intensities of the best data set were multiplied or divided by $\sqrt{E}$. The statistics in figure \ref{fig:paramerrors} summarize these two data sets.  By design, neither $R_\mathrm{P}$ not $R_\mathrm{S}$ should be sensitive to such a ``slow'' intensity variation, and the results confirm this.%
\footnote{Although $R_\mathrm{P}$ and $R_\mathrm{S}$ have very similar (and very weak) sensitivity towards a slow variation of the intensity scale with energy, figure \ref{fig:paramerrors} shows that the parameters obtained with $R_\mathrm{P}$ are much closer to the reference than those using $R_\mathrm{S}$ for optimization. We have checked that this result is not a statistical fluke. A closer analysis indicates that $R_\mathrm{P}$ locks into a sharp minimum of the rough R-factor hypersurface and this minimum remains stationary upon the slow variation of the intensity scale. This is not unexpected since the cusps of $Y_\mathrm{P}(E)$ (and therefore, the local roughness of the R-factor hypersurface) are hardly affected by the slow intensity variations}
One might expect that $R_\mathrm{ZJ}$ should be very sensitive to an energy-dependent intensity factor, because it uses an energy-independent intensity scale factor $c$ between experiment and calculation, see equation (\ref{eq:RZJ}). Nevertheless, $R_\mathrm{ZJ}$ steers the optimization to reasonable parameter values (though not as good as the other $R$ factors).

\textit{Intensity offset}---A constant of $0.005 I_\mathrm{max}$ was added to all intensities, where $I_\mathrm{max}$ is the overall maximum of the set of $I(E)$ curves. Note that this intensity offset corresponds to hardly more than one least-significant bit in legacy 8-bit image acquisition systems, if the brightest beams just reach the saturation limit.%
\footnote{In 8-bit images, the least significant bit corresponds to $1/255 \approx 0.004$ of the maximum intensity. The data used here \cite{kraushofer_viperleed_2025} were obtained with a 12-bit image acquisition system; binning of adjacent pixels has further increased the dynamic range.}
Nevertheless, for a few low-intensity beams in our data set \cite{kraushofer_viperleed_2025}, the offset is higher than the average intensity of the $I(E)$ curve. The effect of such an intensity offset on the parameters is comparable for $R_\mathrm{P}$ and $R_\mathrm{S}$. With $R_\mathrm{ZJ}$, the effect of an intensity offset is significantly worse.

\textit{Incomplete data}---These $I(V)$ curves contained only half of the beams in the full data set. (The statistics in the bottom section of figure \ref{fig:paramerrors} are based on two such partial data sets, one half of the beams and the other half, and the statistics summarize the results for the two.) Optimization with incomplete data sets is meant as a test for how well the $R$ factor makes use of the information in the $I(E)$ data. Ignoring important aspects of the data would lead to overfitting and therefore a poor agreement with the fit using the full data set. At least for the data set used here, the results of the three $R$ factors are comparable; $R_\mathrm{ZJ}$ performs slightly worse than the other $R$ factors. A closer examination of the individual parameter values shows that $R_\mathrm{P}$ and $R_\mathrm{S}$ yield very similar values, substantially deviating from those of $R_\mathrm{ZJ}$ (not shown). This is not unexpected since $R_\mathrm{P}$ and $R_\mathrm{S}$ share the same concept.  It should be noted that, like all the test data used here, also this data set is a rather moderate deterioration of the best data: The redundancy factor \cite{heinz_electron_2013,fauster_surface_2020} for these incomplete data sets is  $\rho\approx 10$, so one should not expect severe overfitting (total energy span ${\approx}5500$\,eV, $|V_\mathrm{0i}|=4.44$\,eV, 31 parameters including the offset of the energy axis or $V_\mathrm{0r}$).

Summarizing all results in figure \ref{fig:paramerrors}, the fit results are significantly more sensitive to imperfections of the data when using $R_\mathrm{ZJ}$ compared with the other $R$ factors. This may be related to a comparably low sensitivity of $R_\mathrm{ZJ}$ to the fit parameters \cite{tear_comparison_1981}, which is also seen as a lower ratio between the curvature and minimum $R$ factor in figure \ref{fig:parabolas}a. The newly introduced $Y_\mathrm{S}$ wins the comparison with $R_\mathrm{P}$, but the individual results should be seen with care since the noisiness of $R_\mathrm{P}$ also adds noise to the values obtained with $R_\mathrm{P}$.

\subsection{\label{sec:performance}Impact on computational performance}

The choice of the $R$ factor has a major impact on the computational performance of the minimization procedure. The computing time is usually dominated by the calculation of the $I(E)$ curves; the time for the R-factor evaluation is comparably low and the increased complexity of $R_\mathrm{S}$ vs.\ $R_\mathrm{P}$ has no substantial impact on the overall performance. The number of trial structures required to find the minimum is much more important. The noise of $R_\mathrm{P}$ makes gradient-based minimization less efficient since a noisy gradient can steer the minimization in the wrong direction. In the minimization runs performed for figure \ref{fig:paramerrors}, the average computing time required for optimization with $R_\mathrm{P}$ as an objective function is 30--50\% higher than when optimizing $R_\mathrm{S}$ or $R_\mathrm{ZJ}$. While this is not particularly bad, the main problem comes from the fact that $R_\mathrm{P}$-based minimization with efficient gradient-based methods will often fail to find the minimum of $R_\mathrm{P}$; thus one needs many search runs. Due to many local minima in close proximity to each other (figure \ref{fig:contours}a), finding the global minimum is even more difficult than just finding a ``reasonable'' minimum well within the error bars of the parameters. For the reference structure of figure \ref{fig:paramerrors}, when using $R_\mathrm{P}$, the best minimum was found in only one of 150 optimization runs. Although the average deviations of the parameters at the lowest minima are only in the order of one tenth of their error bars, some minimization runs (and also the deviations of some parameters in ``good'' runs) end much further out. This is an unsatisfying situation. In principle, one could select minimization methods that are more robust with respect to the $R_\mathrm{P}$-induced noise, but these require substantially more computing time \cite{imre_jax_2026}. Clearly, a better solution is using $R_\mathrm{S}$ for the optimization. With $R_\mathrm{S}$, more than 98\% of all minimization runs land closer to the minimum than the best one percent of all minimization runs using $R_\mathrm{P}$.

\section{\label{sec:conclusions}Conclusions}

For the structural search by LEED $I(E)$, Pendry's $R$ factor $R_\mathrm{P}$ has many attractive properties, but also suffers from severe shortcomings. We have presented a new $R$ factor, named $R_\mathrm{S}$, that retains the advantages of $R_\mathrm{P}$, but avoids its shortcomings. The differences between $R_\mathrm{S}$ and $R_\mathrm{P}$ are limited to a different handling of the high (or even diverging) values of the logarithmic derivative at the minima of the $I(E)$ curves. The numerical values of $R_\mathrm{S}$ and $R_\mathrm{P}$ are similar, and also the  $R_\mathrm{P}$-based method of estimating the error bars of the fit parameters can be directly transferred to $R_\mathrm{S}$. We show that $R_\mathrm{S}$ avoids the noisiness of $R_\mathrm{P}$, thus it is much more suitable as an objective function for minimization. Based on minimization runs with imperfect experimental data, we show that $R_\mathrm{S}$ is at least as good as $R_\mathrm{P}$ in steering the minimization towards the right structure, whereas the optimization results obtained using $R_\mathrm{ZJ}$ are clearly more sensitive to experimental imperfections.


\funding{This research was funded in part by the Austrian Science Fund (FWF) under project 10.55776/F81, Taming Complexity in Materials Modeling (TACO). The computational results presented have been achieved in part using the Austrian Scientific Computing (ASC) infrastructure. For the purpose of open access, the authors have applied a CC BY public copyright license to any Author Accepted Manuscript version arising from this submission. The authors acknowledge TU Wien Bibliothek for financial support through its Open Access Funding Programme.}


\data{Supplementary file \texttt{ImprovedReliabilityFactorData.zip} contains the computer code to create figures \ref{fig:YPe} and \ref{fig:atMin} as well as the data used for the other figures.}



\bibliographystyle{iopart-num-mod}
\bibliography{Modified_RPendry}

\end{document}